# Tuning the valley and chiral quantum state of Dirac electrons in van der Waals heterostructures

J. R. Wallbank[1,2], D. Ghazaryan[2], A. Misra[2], Y. Cao[3], J. S. Tu[3], B. A. Piot[4], M. Potemski[4], S. Pezzini[5], S. Wiedmann[5], U. Zeitler[5], T. L. M. Lane[1,2], S. V. Morozov[2,6,7], M. T. Greenaway[8], L. Eaves[2,8], A. K. Geim[3], V. I. Fal'ko[1,2*], K. S. Novoselov[1,2*], A. Mishchenko[1,2*]

[1]*National Graphene Institute, University of Manchester, Manchester, M13 9PL, UK*

[2]*School of Physics and Astronomy, University of Manchester, Manchester, M13 9PL, UK*

[3]*Centre for Mesoscience and Nanotechnology, University of Manchester, M13 9PL, UK*

[4] *Laboratoire National des Champs Magnétiques Intenses, LNCMI-CNRS-UGA-UPS-INSA-EMFL, 25, avenue des Martyrs, 38042 Grenoble, France*

[5] *High Field Magnet Laboratory (HFML-EMFL) and Institute of Molecules and Materials, Radboud University, Nijmegen, 6525 ED, the Netherlands*

[6]*Institute of Microelectronics Technology and High Purity Materials, RAS, Chernogolovka 142432, Russia*

[7]*National University of Science and Technology "MISiS", 119049, Leninsky pr. 4, Moscow, Russia*

[8]*School of Physics and Astronomy, University of Nottingham NG7 2RD UK*

\* To whom correspondence should be directed: vladimir.falko@manchester.ac.uk , kostya@manchester.ac.uk  and artem.mishchenko@manchester.ac.uk

**Chirality is a fundamental property of electrons with the relativistic spectrum found in graphene and topological insulators. It plays a crucial role in relativistic phenomena, such as Klein tunneling, but it is difficult to visualize directly. Here we report the direct observation and manipulation of chirality and pseudospin polarization in the tunneling of electrons between two almost perfectly aligned graphene crystals. We use a strong in-plane magnetic field as a tool to resolve the contributions of the chiral electronic states that have a phase difference between the two components of their vector wavefunction. Our experiments not only shed light on chirality, but also demonstrate a technique for preparing graphene's Dirac electrons in a particular quantum chiral state in a selected valley.**

The chiral properties of Dirac electrons in monolayer graphene(*1-3*) (and the Berry phase π associated with them) have been used to explain Klein tunneling, the absence of backscattering in graphene p-n junctions(*4-6*), specific features in weak localization(*7, 8*), a peculiar Landau level spectrum (in which one level is pinned exactly at the Dirac point leading to the 'half-integer' quantum Hall effect)(*1, 2*), and valley-selection of the interband transitions excited by polarized light(*9*). Chirality is determined by the relative phase in the two-component wavefunction of the Dirac quasi-particles, which arises from the sublattice composition in graphene(*3*) and from spin states in topological insulators(*10*). Such a two-component wavefunction is typically described in terms of a specific vector - the pseudospin – which, for chiral particles, is locked to their direction of motion. However, it has proved difficult to image directly the chirality and the pseudospin polarization in electrical or optical measurements. To date the phase shift in the sublattice composition of the electron states in graphene has been detectable only by angle resolved photoemission spectroscopy (ARPES)(*11-13*).

Here we report an alternative technique for pseudospin and chirality detection that is based on tunneling of electrons in van der Waals (vdW) heterostructures(*14, 15*) in which graphene (Gr) and hexagonal boron nitride (hBN) are stacked in a multilayer structure (Fig. 1A). In these devices, it has been shown that the exceptionally high quality of graphene, provided by its encapsulation, allows the electrons to tunnel between graphene electrodes with conservation of in-plane momentum (*16, 17*), making one graphene electrode a bias voltage ($V_b$)-tunable spectrometer for electrons emitted by the other. However, the usual tunneling formalism, which does not take into account the interference between the two components of the wavefunction of the tunneling electrons, fails in the case of chiral quasiparticles. Here, we show that, the tunneling current-voltage characteristics $I(V_b,B)$ in the presence of an in-plane magnetic field $B$ *(18)*, essentially depend on the pseudospin orientation and enable us to detect the valley sublattice structure determined by the relative phase between the two sublattice components of the Dirac spinor vector wavefunction of electrons in graphene.

A small misalignment angle between the crystalline lattices of the two graphene flakes (Fig. 1A) causes specific parts of $I(V_b,B)$ to arise from the tunneling of electrons in specific regions of momentum space. Furthermore, for the case of chiral electrons, different states in momentum space (and thus with specific pseudospin orientation) have different tunneling probability – depending on whether the interference between the two components of electron wavefunction is constructive (Fig. 1B-D) or destructive (Fig. 1F-H) as electrons tunnel out of the emitting graphene layer. A particular state can be chosen with the help of a magnetic field, applied perpendicular to the current. This provides the electrons with a tunable momentum boost as they traverse the barrier (as shown previously in studies of vertical transport in III-V semiconductor heterostructures(*19, 20*)). For graphene, by rotating the magnetic field in the plane of the Gr/hBN/Gr device, we are able to resolve the contributions to the measured differential conductance, $G = \partial I/\partial V_b$, arising from electrons with clearly identifiable momenta in a given valley of graphene's band structure, and hence detect the features related to the sublattice composition of the electronic wavefunctions.

In the series of vdW heterostructures studied here, a tunnel barrier of hBN separates a graphene monolayer from either a monolayer or a bilayer of graphene. During preparation, we ensure that the crystallographic orientations of the two graphene electrodes are closely aligned, Fig. 1A, by aligning the edges of the flakes in the transfer procedure (for details, see (*21, 22*)). The devices are placed on the oxidized surface of a doped silicon substrate which forms an insulated back gate electrode, as in previously-reported graphene tunneling transistors(*16, 17, 23*). The two types of multilayer stack are thus of the form Si/SiO$_2$/hBN/Gr/N-hBN/(B)Gr/hBN, where (B)Gr denotes (bilayer) graphene and N-hBN denotes N layers of hBN. The typical active areas of our devices are between 10 and 100μm².

Typical plots of *G* versus $V_b$, and back gate voltage, $V_g$, for the Gr/3hBN/Gr and Gr/5hBN/BGr devices are presented in Fig. 2 (see examples of other aligned devices in (*22*)). In these aligned devices a number of resonant features in the tunneling $I(V_b)$ characteristics are observed, such as when the Fermi level in one layer coincides with the lowest energy at which the band dispersion curves of the two layers intersect (Fig. 2A). Schematic representations of some of these resonant alignment conditions are shown in Fig. 2 with more details given in (*22*). These resonances are absent in the devices in which the graphene electrodes are strongly misaligned; for these devices momentum conservation is satisfied by elastic scattering and/or phonon emission (*22, 24-26*).

There are qualitative differences in the tunneling conductance plots of the Gr/hBN/Gr and Gr/hBN/BGr devices mainly thanks to (i) the difference in the density of states (DoS) between graphene (DoS is linear with energy) and bilayer graphene (DoS is independent of energy) and (ii) the presence of the second subband in bilayer graphene. Thus, the difference in DoS leads to most features in the conductivity plot for Gr/hBN/Gr having a square root dependence in the $V_b$-$V_g$ plane, whereas some of these features are linear for the Gr/hBN/BGr devices (compare blue dashed lines in Fig. 2D and 2H). Also, the presence of the second subband in the bilayer roughly doubles the number of the observed resonances.

To gain further insight, we computed the tunnel conductance using a previously developed model of the device electrostatics(*16, 23*) and a chiral tunneling formalism(*27*). The only free parameters in the model are the relative angle between the crystallographic directions of the graphene flakes and the energy broadening of their plane wave states. By comparing the experimental and calculated results in Fig. 2, we can extract the relative orientation angle between the two graphene flakes, which we find to cover the range 0.5° to 3° for the group of devices that we studied.

A strong magnetic field, ~30T, applied parallel to the graphene layers gives rise to additional fine structure in $G$ (Fig. 3G), which is strongly dependent on the angle between $B$ and the principal axes of the graphene crystals. This is best revealed in the $G' \equiv \partial G/\partial V_b$ plots (Fig. 3, H and K), with further examples presented in (*22*). The angular dependence of these features corresponds to six intertwined sinusoids.

The origin of this fine structure is explained in Fig. 3, A-F. In zero magnetic field all six corners of the Brillouin zone (BZ) experience resonance conditions simultaneously, leading to a single resonance peak (Fig. 3A-C, where an example is given for the resonant alignment, similar to the "touch" depicted in Fig. 2A). In a finite magnetic field, the tunneling electrons experience a Lorentz force, and gain an in-plane momentum boost, given by

$$\Delta \vec{p} = ed\hat{z} \times \vec{B}.$$

Here, $\hat{z}$ is a unit vector in the tunneling direction, $e$ is the electron charge and $d$ is the thickness of hBN tunneling barrier. In Fig. 3, D and E this is represented as a relative shift of the two BZ, additional to the rotation arising from the small angular misalignment of the two graphene layers. Hence, depending on the orientation of magnetic field, the resonant conditions for the six corners of the BZ are fulfilled at six slightly different voltages, leading to the splitting of the resonance peak into six individual peaks. The $V_b$ value required for resonance at each particular corner is a sinusoidal function of the angle α between $B$ and the "armchair" direction of the graphene lattice: $V_b(B,\alpha,j) \approx V_b(B=0) + \Delta V(B,\alpha)\sin(\alpha+j\pi/3)$, where j=0,1..5 is the index of a particular *BZ* corner (as shown by grayscale-colored lines in Fig 3H and K). Our theoretical model provides a very good fit to our experimental results (Fig. 3, I and L).

The intensities of these resonances also depend on α, so only half of the period of the sinusoid is visible (Fig. 3, H,I,K,L). This is particularly obvious in Fig. 3, K and L for the resonance between $V_b$=0V and 0.25V. This asymmetry arises from the electronic chirality of graphene. The electron wavefunction is a vector with two components, $\psi_A$ and $\psi_B$, representing the probability of finding the electron on the two sublattices, *A* and *B,* of the honeycomb lattice. Chirality is the specific property of the relative phase $\varphi$ between the wavefunction components, which is locked to the direction of the electron's momentum, $\vec{p} = p(\cos\vartheta, \sin\vartheta)$, counted from the nearest BZ corner. For a monolayer $\varphi = \vartheta$, for gapless bilayer graphene it would be $\varphi = 2\vartheta$ for *A/B'* sublattices supporting low-energy bands.

Electron tunneling from one graphene layer to another requires a correlation between the two components of the wavefunction in both layers. In effect, this projects the wavefunction in the graphene emitter and collector onto evanescent waves in the barrier space between two 2D-crystals (in case of BLG we project only the wavefunctions located on the layer closest to the hBN tunneling barrier). The projected states are composed of two sublattice components in the emitter and collector. As a result, momentum-dependent constructive ($\varphi_{e(c)}=0$) or destructive ($\varphi_{e(c)}=\pi$) interference between sublattice components is governed by $|\psi_A + \psi_B|^2 \propto 1 + cos\varphi_{e(c)}$, for the states both in emitter ($\varphi_e$) and collector ($\varphi_c$) and manifests itself in the tunneling characteristics $I(V_b)$. Because the magnetic field selects the pairs of particular plane wave states probed by tunneling at a particular gate or bias voltage (Fig. 4, A and B), the measured asymmetry provides a direct visualization of the pseudospin polarization of the Dirac fermions.

In the presence of the magnetic field, each resonance peak represents tunneling from a particular corner of the BZ. This allows us to inject electrons with a particular valley polarization, and from a selected corner of the BZ. We use the experimental parameters to calculate the amount of polarization achieved in our experiment (Fig. 3, J and M), and estimate that the valley polarization, $P = (I_K - I_{K'})/(I_K + I_{K'})$ (where $I_K(I_{K'})$ is the current injected into the K(K') valley) can be as high as 30% (40%) for the particular Gr/3hBN/Gr (Gr/5hBN/BGr) devices. The main limit to the degree of polarization is the energy broadening of states at the Fermi levels caused by inelastic tunneling processes. However, even for the current level of disorder, utilizing the resonances at around $V_b \approx 0V$ (e.g. resonances marked by yellow dashed lines on Fig. 2D at $V_g > 50V$) which maximizes the number of states participating in tunneling and sensitive to magnetic field, a polarization close to 75% could be achieved(*22*). Using devices with smaller misalignment between the graphene electrodes (of order of 0.2°, now within the reach of the current technology (*22*)) valley polarization close to 100% is possible(*22*).

The same mechanism can also be used to select electrons with a particular pseudospin polarization. In Fig. 4, C-R we present results of a calculation of the contribution of different electronic states in *k*-space to the tunnel current for the Gr/3hBN/Gr (Fig. 4C-I) and Gr/5hBN/BGr (Fig. 4, J-R) devices. We choose the position of the Fermi levels in the emitter and collector to be very close to a resonance at $B$=0T. Then, for certain directions of $B$, the resonant conditions are achieved only in one valley and for only a very narrow distribution in $k$-space (Fig. 4, G-I). Tunneling of the electrons from other parts of $k$-space is prohibited either because they are off-resonance or because of the pseudospin selection rule. Alternatively, for the Gr/5hBN/BGr device and exploiting the difference in curvature of monolayer and bilayer electronic bands, we can choose the overlap between the bands in such a way that the magnetic field reduces the overlap in one valley and increases it for the other (Fig. 4, M-R). In this case momentum conservation at $B$=0T is fulfilled for the states marked by white dashed lines, Fig. 4O. However only one of those lines contributes to tunneling, due to pseudospin interference (Fig. 4, M and N).

Our technique, which enables tunneling of valley-polarized electrons in monolayer and bilayer graphene, also allows one to inject selectively carriers propagating in the same direction and to probe pseudospin-polarized quasiparticles. In principle, the technique can be extended to tunneling devices in which surface states of topological insulators are used as electrodes; then all-electrical injection of spin-polarized current(*28*) using non-invasive tunneling contacts could reveal a number of exciting phenomena(*29-31*).

1.  K. S. Novoselov *et al.*, *Nature* **438**, 197-200 (2005).
2.  Y. B. Zhang, Y. W. Tan, H. L. Stormer, P. Kim, *Nature* **438**, 201-204 (2005).
3.  A. H. Castro Neto, F. Guinea, N. M. R. Peres, K. S. Novoselov, A. K. Geim, *Rev. Mod. Phys.* **81**, 109-162 (2009).
4.  M. I. Katsnelson, K. S. Novoselov, A. K. Geim, *Nat. Phys.* **2**, 620-625 (2006).
5.  M. I. Katsnelson, *EPJ B* **51**, 157-160 (2006).
6.  A. F. Young, P. Kim, *Nat. Phys.* **5**, 222-226 (2009).
7.  F. V. Tikhonenko, D. W. Horsell, R. V. Gorbachev, A. K. Savchenko, *Phys. Rev. Lett.* **100**, 056802 (2008).

8. E. McCann *et al.*, *Phys. Rev. Lett.* **97**, 146805 (2006).
9. M. L. Sadowski, G. Martinez, M. Potemski, C. Berger, W. A. de Heer, *Phys. Rev. Lett.* **97**, 266405 (2006).
10. X. L. Qi, S. C. Zhang, *Rev. Mod. Phys.* **83**, 1057-1110 (2011).
11. S. Y. Zhou *et al.*, *Nat. Phys.* **2**, 595-599 (2006).
12. A. Bostwick, T. Ohta, T. Seyller, K. Horn, E. Rotenberg, *Nat. Phys.* **3**, 36-40 (2007).
13. M. Mucha-Kruczynski *et al.*, *Phys. Rev. B* **77**, 195403 (2008).
14. A. K. Geim, I. V. Grigorieva, *Nature* **499**, 419-425 (2013).
15. G. H. Lee *et al.*, *Appl. Phys. Lett.* **99**, 243114 (2011).
16. A. Mishchenko *et al.*, *Nat. Nanotechnol.* **9**, 808-813 (2014).
17. B. Fallahazad *et al.*, *Nano Lett.* **15**, 428-433 (2015).
18. L. Pratley, U. Zulicke, *Phys. Rev. B* **88**, 245412 (2013).
19. R. K. Hayden *et al.*, *Phys. Rev. Lett.* **66**, 1749-1752 (1991).
20. V. I. Falko, S. V. Meshkov, *Semicond. Sci. Technol.* **6**, 196-200 (1991).
21. A. V. Kretinin *et al.*, *Nano Lett.* **14**, 3270-3276 (2014).
22. *Supplementary Materials*.
23. L. Britnell *et al.*, *Science* **335**, 947-950 (2012).
24. F. Amet *et al.*, *Phys. Rev. B* **85**, 073405 (2012).
25. S. Jung *et al.*, *Sci. Rep.* **5**, 16642 (2015).
26. E. E. Vdovin *et al.*, *Phys. Rev. Lett.* **116**, 186603 (2016).
27. T. L. M. Lane, J. R. Wallbank, V. I. Fal'ko, *Appl. Phys. Lett.* **107**, 203506 (2015).
28. C. H. Li *et al.*, *Nat. Nanotechnol.* **9**, 218-224 (2014).
29. D. Culcer, E. H. Hwang, T. D. Stanescu, S. Das Sarma, *Phys. Rev. B* **82**, 155457 (2010).
30. O. V. Yazyev, J. E. Moore, S. G. Louie, *Phys. Rev. Lett.* **105**, 266806 (2010).
31. D. Pesin, A. H. MacDonald, *Nat. Mater.* **11**, 409-416 (2012).
32. L. A. Ponomarenko *et al.*, *Nat. Phys.* **7**, 958–961 (2011).
33. L. Wang *et al.*, *Science* **342**, 614-617 (2013).
34. J. R. Wallbank, *Electronic properties of graphene heterostructures with hexagonal crystals*. Springer PhD Thesis Series (Springer, 2014).
35. R. M. Feenstra, D. Jena, G. Gu, *J. Appl. Phys.* **111**, 043711 (2012).

36. F. T. Vasko, *Phys. Rev. B* **87**, 075424 (2013).

37. S. C. de la Barrera, Q. Gao, R. M. Feenstra, *J. Vac. Sci. Technol. B* **32**, 04E101 (2014).

38. L. Brey, *Physical Review Applied* **2**, 014003 (2014).

39. K. A. Guerrero-Becerra, A. Tomadin, M. Polini, *Phys. Rev. B* **93**, 125417 (2016).

40. E. McCann, *Phys. Rev. B* **74**, 161403 (2006).

41. E. McCann, M. Koshino, *Rep. Prog. Phys.* **76**, 056503 (2013).

42. J. R. Schrieffer, Scalapin.Dj, J. W. Wilkins, *Phys. Rev. Lett.* **10**, 336 (1963).

43. G. D. Mahan, *Many particle physics.* (Plenum Press, ed. 2nd edition, 1990).

44. E. H. Hwang, B. Y. K. Hu, S. Das Sarma, *Phys. Rev. B* **76**, 115434 (2007).

45. E. H. Hwang, S. Das Sarma, *Phys. Rev. B* **75**, 205418 (2007).

46. D. M. Basko, S. Piscanec, A. C. Ferrari, *Phys. Rev. B* **80**, 165413 (2009).

47. M. Lazzeri, C. Attaccalite, L. Wirtz, F. Mauri, *Phys. Rev. B* **78**, 081406 (2008).

48. S. Piscanec, M. Lazzeri, F. Mauri, A. C. Ferrari, J. Robertson, *Phys. Rev. Lett.* **93**, 185503 (2004).

49. J. W. McClure, *Phys. Rev.* **104**, 666-671 (1956).

This work was supported by the EU FP7 Graphene Flagship Project 604391, ERC Synergy Grant, Hetero2D, EPSRC (Towards Engineering Grand Challenges and Fellowship programs), the Royal Society, US Army Research Office, US Navy Research Office and US Airforce Research Office. M.T.G acknowledges support from the Leverhulme Trust. A.M. acknowledges support of EPSRC Early Career Fellowship EP/N007131/1. S.V.M. was supported by NUST "MISiS" (grant K1-2015-046) and RFBR (15-02-01221 and 14-02-00792). Measurements in high magnetic field were supported by HFML-RU/FOM and LNCMI-CNRS, members of the European Magnetic Field Laboratory (EMFL) and by EPSRC (UK) via its membership to the EMFL (grant no. EP/N01085X/1).

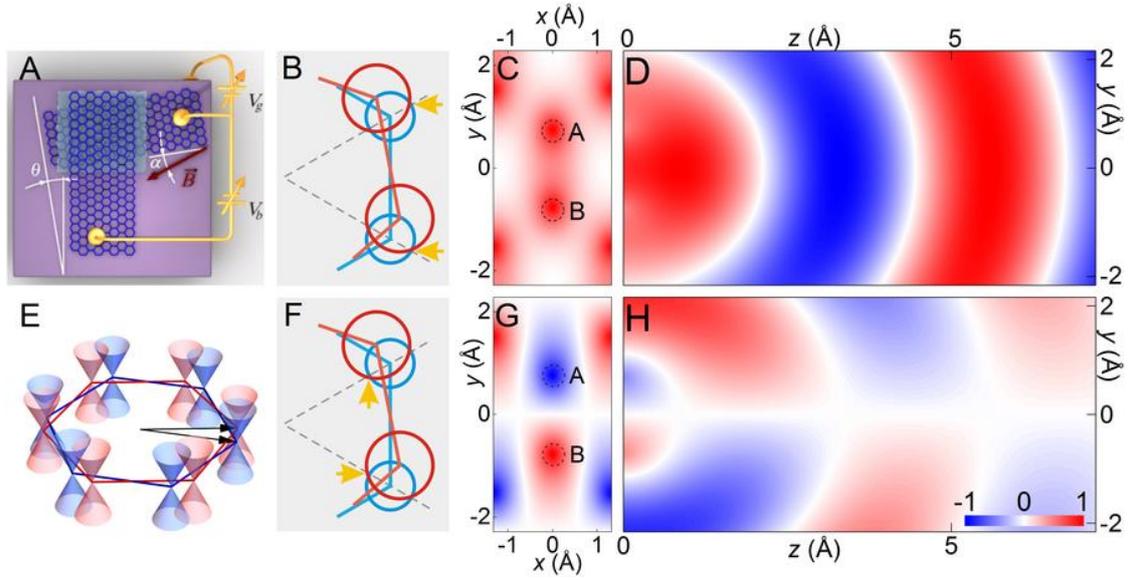

**Fig.1. Device schematics, band structure and chiral composition of the wavefunction.** (**A**) Schematic diagram of Si/SiO$_2$/Gr/hBN/Gr device (dark blue hexagonal layers are graphene electrodes, light blue – hBN, purple – Si/SiO$_2$ back gate). Gate voltage $V_g$ is applied between the bottom graphene and Si substrate. (**B**) Two corners of the BZ (full BZ is shown in (E)) schematically demonstrating the Fermi surfaces for emitter (blue circles) and collector (red circles). Yellow arrows mark the states in the emitter when the components of the wavefunction on the two sublattices are in-phase (shown in (C)). (**C**) Real-space distribution of the real part of the wavefunction on the A and B sublattices. The two components of the wavefunction are in-phase. (**D**) Schematic representation of the interference of the two-component wavefunction of graphene electron at the given distance above the graphene layer when the electron is taken away from graphene. The original electron state at $z=0$ is as in (C). (**E**) Small-angle rotational misalignment between the two graphene crystals leads to a small momentum mismatch between the two band structures in the reciprocal space. (**F**) Same as (B). Yellow arrows mark the states in the emitter when the components of the wavefunction on the two sublattices are out-of-phase (shown in (G)). (**G**) Real-space distribution of the real part of the wavefunction on the A and B sublattices. The two components of the wavefunction are out-of-phase. (**H**) Same as in (D), but when the original electron state at $z=0$ is as in (G). Colour scales for (C,D,G,H) are the same.

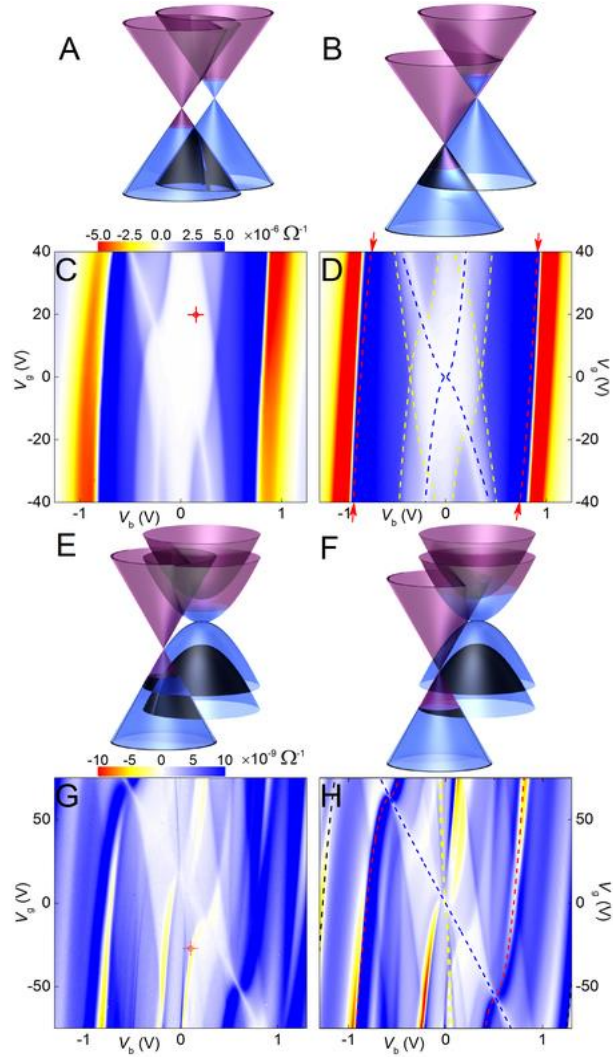

**Fig. 2. Tunneling characteristics of our devices.** (**A, B**) Relative position of the bands and the Fermi levels in the two graphene electrodes rotated by a small angle with respect to each other for conditions highlighted by yellow (A) and red (B) dashed lines in (D). Red dashed lines in (D) are also marked by red arrows for clarity. (**C,D**) Experimental (C) and simulated (D) tunneling characteristics for a Gr/3hBN/Gr device with the graphene electrodes misaligned by 1.8°. The red cross in (C) marks the $V_b$ and $V_g$ used for the calculations of chirality polarization in Fig. 4, C-I. The blue lines in (D,H) mark the conditions when the Fermi level in one of the electrodes passes through the Dirac point where the DoS is zero which leads to the suppression of tunneling conductance. (**E,F**) Relative position of the bands and the Fermi levels in the Gr and BGr electrodes rotated by a small angle with respect to each other for the resonant conditions highlighted by red (E) and black (F) dashed lines in (H). (E) – the low energy subband in the valence band in bilayer graphene touches the graphene cone. (F) – the higher energy subband in the valence band in bilayer graphene touches the graphene cone. (**G, H**) Experimental (G) and simulated (H) tunneling

characteristics for a Gr/5hBN/BGr device with the graphene electrodes misaligned by 0.5° (a small part of the sample is misaligned by 3°, which explains some of the weaker features). The yellow dashed lines mark the resonance when the Fermi level in monolayer graphene touches the bottom (top) of the conduction (valence) band in the bilayer graphene. The red cross in (G) marks the $V_b$ and $V_g$ used for the calculations of chirality polarization in Fig. 4(M-R).

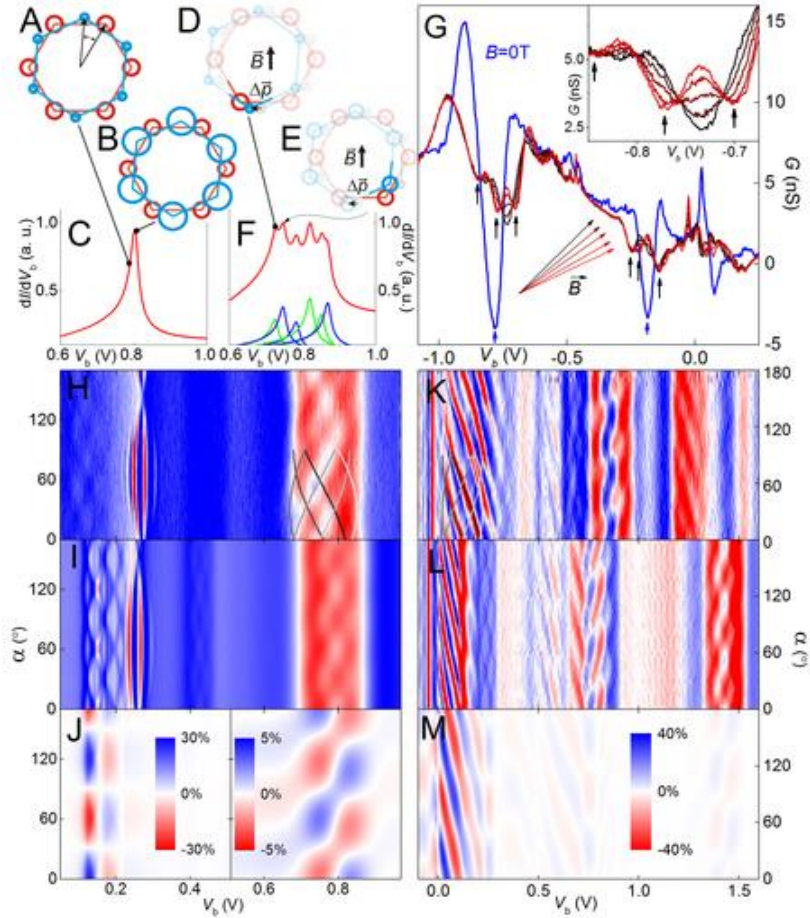

**Fig. 3. Magnetotunneling characteristics of studied devices.** (**A**) Schematic representation of the BZ for the emitter (blue) and collector (red) graphene electrodes rotated by a small angle with respect to each other. Circles represent the Fermi surfaces in the two graphene layers. (**B**) As (A) but with increased Fermi level in the emitter, which induces resonant condition of the type presented in Fig. 2A for all 6 corners of the BZ simultaneously. (**C**) Resonance in $dI/dV_b$ corresponding to (B). (**D**) As (A), but with $B$ applied parallel to graphene layers. The Lorentz force leads to an additional momentum acquired by electrons when tunneling from the emitter to the collector, which can be represented by a relative shift of the two BZ by the vector $\Delta\vec{p}$ (grey – the BZ in $B$=0, blue – in finite $B$). This can bring different corners of the BZ into resonance, depending on α. (**E**) As (D) but at different doping (increased Fermi level in the emitter), which brings a different corner of the BZ into resonance. (**F**) The resonant peak in d$I$/d$V_b$ splits into 6 peaks in finite magnetic field (red curve), each corresponding to a resonance that occurs in each corner of the BZ (green and blue curves for K and K' valleys respectively). Examples for particular resonant conditions for two corners of the BZ are shown in (D) and (E). (**G**) Conductance of the Gr/5hBN/BGr device at $V_g$=-45V for $B$ = 0T (blue) and 30T for α increasing from 0° in 5° steps (black to red). Note that some minima (marked by short blue arrows) are split by the magnetic field (black arrows), see enlarged inset. (**H,I**). d$G$/d$V_b$ vs $V_b$ and $\alpha$ for the Gr/3hBN/Gr device with $V_g$=20V. (**K,L**). d$G$/d$V_b$ vs $V_b$ and $\alpha$ for Gr/5hBN/BGr device with $V_g$=60V. Panels (H) and (K) – experimental data; (I) and (L) – theory. The six black, grey and white lines in (H) and (K) are guides to the eye and mark the position of the resonances for the six corners of the BZ. (**J,M**) Calculated valley polarisation of tunneling current for Gr/3hBN/Gr (J) and Gr/5hBN/BGr (M) devices. Note different colour scale for $V_b$<0.5V and $V_b$>0.5V in (J).

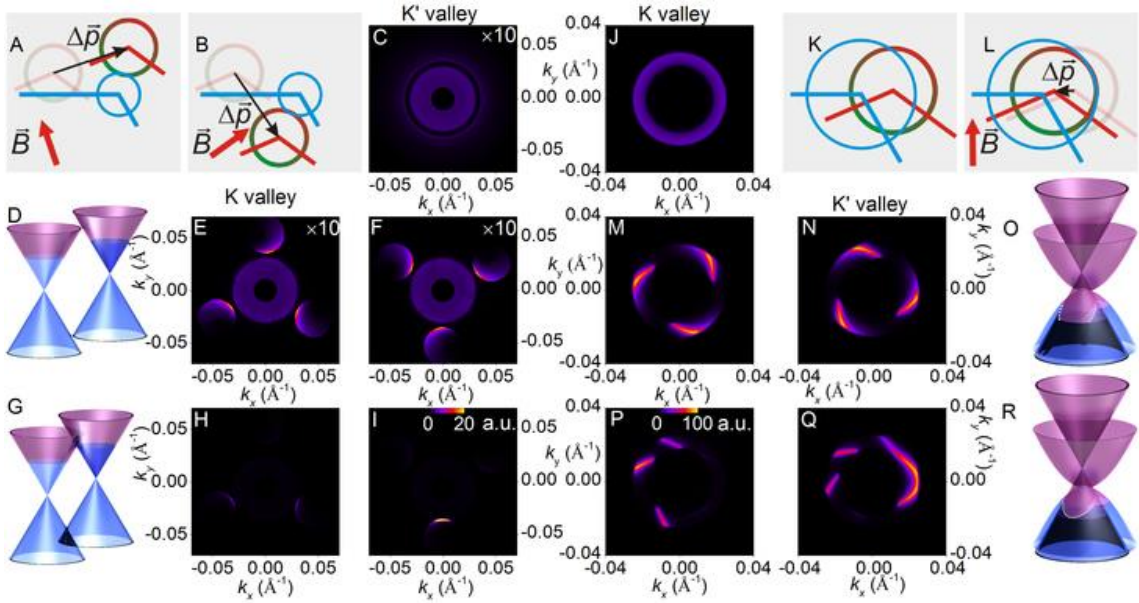

**FIG. 4. Calculated tunneling of chirality-polarized electrons.** (**A**) Schematic representation of a resonant condition for tunneling between two graphene electrodes in a magnetic field. The gradient coloring of the Fermi surface in the collector represents a particular phase difference φ between the wavefunction components (green: φ=π, red: φ=0). Similar coloring for the emitter is omitted for simplicity.

(**B**) As (A) but for a different orientation of magnetic field. States with different φ than in (A) are now in resonance. (**C**) Contribution of different states in the K'-valley of the graphene emitter to the current for the Gr/3hBN/Gr device with $V_b$=0.13V, $V_g$=20V and $B$=0T (marked by a red cross in Fig. 2C) for the case of infinitely short elastic scattering time (inelastic tunneling). States with all orientations of momentum contribute equally to tunneling. The situation in the K valley is the same, and the colour scale is enhanced by ×10 compared to (H,I). (**D**) Relative position of the Fermi energies in the two graphene electrodes for $V_b$=0.13V and $V_g$=20V at $B$=0T. (**E-F**) The contribution of different states in the K-valley or K'-valley of the graphene emitter to the current for the conditions presented in (D). No resonant conditions are achieved, so the small current observed in (E, F) is a consequence of the energy broadening of states (colour scale enhanced by ×10 compared to (H,I)). (**G-I**) similar to (D-F) except with $B$=30T. Resonant conditions are now achieved in the K'-valley only (depicted in G), which is reflected by the current distribution in (I). (**J-R**) Similar to (A-I) except for the Gr/5hBN/BGr device at $V_b$=0.045V and $V_g$=-30V (marked by a red cross in Fig. 2G), with either $B$=0T (J,K,M,N,O) or $B$=30T (L,P,Q,R). There is no enhancement of the color scale between (J,M,N) and (P,Q). The resonant states in (O,R) are highlighted in using dashed white lines. Only one of the two lines of resonant states in (O) produces a visible contribution to the tunneling current (M,N). This can be traced using the phase information in (K): for the lower intersection of the two Fermi lines the interference between wavefunction components is destructive (green).

# Supplementary Materials:
## Tuning the valley and chiral quantum state of Dirac electrons in van der Waals heterostructures

J. R. Wallbank, D. Ghazaryan, A. Misra, Y. Cao, J.S. Tu, B. A. Piot, M. Potemski,
S. Pezzini, S. Wiedmann, U. Zeitler, T. L. M. Lane, S. V. Morozov, M. T. Greenaway,
L. Eaves, A. K. Geim, V. I. Fal'ko*, K. S. Novoselov*, A. Mishchenko*

## 1 Device fabrication

Aligned tunnel transistors were fabricated using the standard dry transfer/peel off technique (*16*, *32*, *33*). Optical micrographs showing the device fabrication steps are displayed in Fig. S1. Relatively thick substrate hBN ($\approx 30\,\mathrm{nm}$) was first exfoliated on Ar+$O_2$ plasma cleaned $SiO_2$/Si substrate. Mono- and bi- layers of graphene obtained by mechanical exfoliation and having well defined long straight edges with multiple of $30°$ were used in device preparation. Monolayer graphene was transferred onto a hBN/$SiO_2$/Si substrate followed by a tunnel hBN ($\approx 3 - 5$ layers) and then mono- or bilayer graphene was transferred on top of the stack. The bottom and top graphene edges were carefully aligned to multiples of $30°$ (while keeping hBN layers misaligned at $\approx 10 - 15°$ to avoid band reconstruction due to moiré superlattice). Since there is an uncertainty whether aligned edges are of the same nature (zigzag or armchair), about $50\%$ of our transistors were aligned and showed resonant behavior. Finally the device was covered by relatively thick ($\approx 30\,\mathrm{nm}$) layer of hBN. To summarize, the stack consists of Si/ $300\,\mathrm{nm}$ $SiO_2$/ $30\,\mathrm{nm}$ hBN/ monolayer Gr/ tunnel hBN/ (B)Gr / $30\,\mathrm{nm}$ hBN. This method of dry peeling of flakes for transfer ensures the minimum PMMA contamination introduced in the layers. The entire stack was annealed at $300°C$ for 3hrs in Ar+$H_2$ atmosphere to remove the bubbles.

Side contacts to the top and bottom graphene layers were made using standard electron beam lithography, reactive ion etching and lift off processes. To make the side contacts, contact regions were defined by electron beam exposure and were etched in a reactive ion etching chamber using $CHF_3$ and $O_2$ mixture. Finally Cr/Au ($3\,\mathrm{nm}/60\,\mathrm{nm}$) contacts were deposited using electron beam induced deposition and lifted off in acetone.

## 2 Transport measurements

In order to perform measurements in in-plane magnetic fields, a device was glued with silver paste to an L-shaped brass holder and then bonded to a chip carrier (LCC). LCC was then mounted in a rotating insert fitted to a $^4$He-cryostat. We use resistive electromagnets (within European Magnetic Field Laboratory facilities in Grenoble and Nijmegen) for measurements in $\approx 30$ Tesla magnetic fields parallel to graphene crystals. Differential conductance was measured employing low-frequency lock-in technique. Small ac excitation (typically $1\,\mathrm{mV}$) was mixed with dc bias via resistive voltage divider and applied to the top graphene layer, ac current from the bottom graphene layer was then amplified by low-noise current preamplifier SR570 (served as virtual ground) and measured with standard lock-in SR830. When necessary, both ac and dc voltage signals were also measured across the tunnel junction.

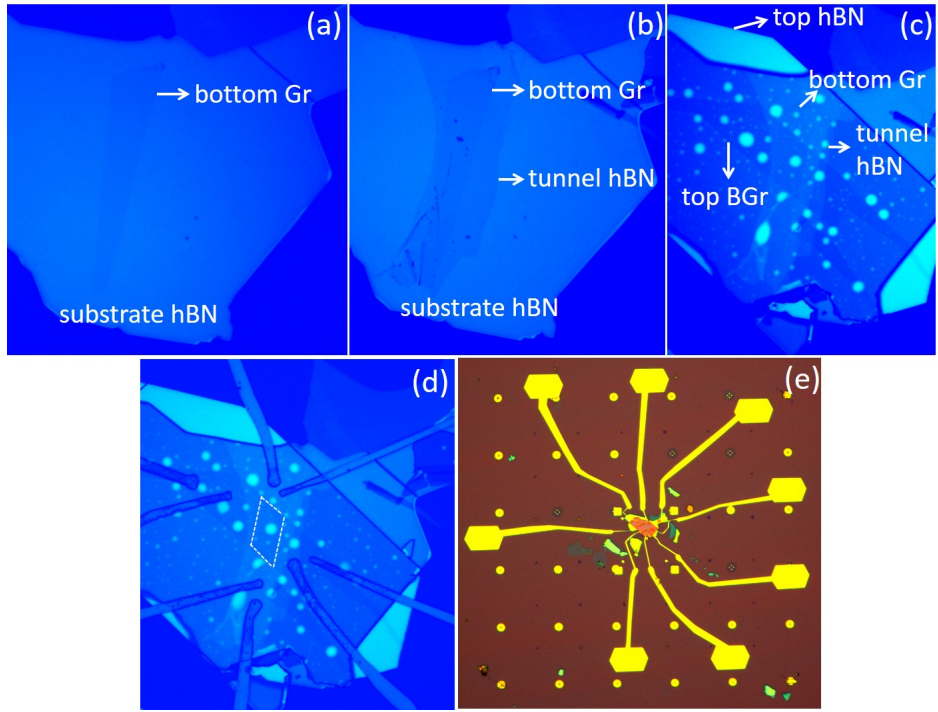

Figure S1: Optical micrographs showing the device fabrication steps: (a) bottom monolayer graphene (bottom Gr) is transferred on top of the substrate hBN; (b) hBN tunnel barrier is transferred on top of the bottom graphene; (c) top bilayer graphene (top BGr) is transferred on the tunnel hBN barrier and then the entire device is encapsulated with top hBN; (d) developed PMMA mask for 1D contacts etching and metalization; (e) Final device. (f) Enlargement of (c) displaying the alignment of crystal axes of the two graphene flakes, resulting in their bandstructures being aligned to within a small angle in K-space. The white spots are bubbles under the top hBN, which do not influence the electronic quality of the major heterostructure.

## 3 Model of the tunneling current in Gr/hBN/Gr and Gr/hBN/BGr devices

Here we describe the calculation of the tunneling for low-energy electrons between graphene layers in Gr/hBN/Gr and Gr/hBN/BGr devices, following models developed in Refs. (*16*, *34*, *27*) and similar theories in Refs. (*35*, *36*, *37*, *38*, *39*). First we present the Hamiltonians used to describe the electronic structure of an isolated Gr/BGr flake, and then, describe the tunneling between them. For an isolated Gr flake on either the bottom/top ($l = B/T$) of a Gr/hBN/Gr stack, or on the bottom ($l = B$) of a Gr/hBN/BGr stack we use the Dirac Hamiltonian,

$$H_l^{\text{Gr}} = v \begin{pmatrix} 0 & \hat{\pi}^\dagger \\ \hat{\pi} & 0 \end{pmatrix}, \tag{S.1}$$

written in a basis of Bloch wavefunctions ($|\phi_{l,\xi}^A\rangle, |\phi_{l,\xi}^B\rangle$) on the $A/B$ graphene sublattices, $v = 6.6\,\text{eV\AA}$ is the Dirac velocity, $\hbar = 1$, and $\hat{\pi} = \xi k_l^x + i k_l^y$ is written using the wavevector $\vec{k}_l = (k_l^x, k_l^y)$ measured from either the $K$ ($\xi = 1$) or $K'$ ($\xi = -1$) Brillouin corner. For the isolated BGr flake on the top ($l = T$) of a Gr/hBN/BGr stack we use (*40*, *41*),

$$H_l^{\text{BGr}} = \begin{pmatrix} \frac{\Delta}{2} & v\hat{\pi}^\dagger & -v_4\hat{\pi}^\dagger & v_3\hat{\pi}^\dagger \\ v\hat{\pi} & \frac{\Delta}{2} & \gamma_1 & -v_4\hat{\pi}^\dagger \\ -v_4\hat{\pi} & \gamma_1 & -\frac{\Delta}{2} & v\hat{\pi}^\dagger \\ v_3\hat{\pi}^\dagger & -v_4\hat{\pi} & v\hat{\pi} & -\frac{\Delta}{2} \end{pmatrix}, \tag{S.2}$$

which is written in a basis of Bloch wavefunctions ($|\phi_{l,\xi}^{A'}\rangle, |\phi_{l,\xi}^{B'}\rangle, |\phi_{l,\xi}^A\rangle, |\phi_{l,\xi}^B\rangle$) on the $A/B$ sublattices of the upper (primed) and lower (un-primed) layers of the BGr flake. Here, $v_3 = 0.67\,\text{eV\AA}$ and $v_4 = 0.32\,\text{eV\AA}$ are produced by the interlayer "skew" hopping of electrons with in the BGr flake, and create trigonal warping and electron-hole asymmetry in the BGr band structure (*41*). Also, the interlayer energy difference, $\Delta$, is produced by electric fields in the device (see section S4), and creates a band gap. To simplify the description of the tunneling between the two Gr/BGr flakes, we take the crystallographic directions of the hBN layer to be highly misaligned from both Gr/BGr layers. Then, any Bragg scattering (processes involving a hBN reciprocal lattice vector) is unable to transfer low-energy electrons between the vicinity of the Brillouin zone corners on the two layers, and would instead result in scattering to high-energy regions of graphene's Brillouin zone ($|\epsilon| \gg |\mu_T|, |\mu_B|$) that do not contribute to tunneling. Because of this, we replace the hBN layer with a structureless "jelly" insulator. Then, we assume that the tunneling matrix element is controlled by the overlap between the tails of the carbon P$_z$ orbitals on the two layers. For a Gr/hBN/BGr device we only account for the tunneling of electrons into the lowest layer of the BGr flake (i.e the layer closest to the Gr), due to the exponential decay of the wavefunction overlaps with the increased distance to the further graphene layer of the BGr. Using this, the matrix element for an electron to tunnel between states with particular band indexes, $s_{B/T}$, and wavevectors, $\vec{k}_{B/T}$, is

$$M_{\vec{k}_T,\vec{k}_B}^{\xi} = \vec{\chi}_{T,\xi}^\dagger \, t \begin{pmatrix} \langle \phi_{T,\xi}^A | \phi_{B,\xi}^A \rangle & \langle \phi_{T,\xi}^A | \phi_{B,\xi}^B \rangle \\ \langle \phi_{T,\xi}^B | \phi_{B,\xi}^A \rangle & \langle \phi_{T,\xi}^B | \phi_{B,\xi}^B \rangle \end{pmatrix} \vec{\chi}_{B,\xi}, \tag{S.3}$$

where $t$ is a hopping integral depending on the thickness of the hBN layer. For Gr, $\vec{\chi}_{l,\xi} = (\chi^A_{l,\xi}, \chi^B_{l,\xi})$ is the normalized eigenvector of Hamiltonian (S.1) describing the tunneling electron, and can be written $\vec{\chi}_{l,\xi} = \frac{1}{\sqrt{2}}(1, \xi s_l e^{i\xi\vartheta_{\vec{k}_l}})$, using $s_l = \pm$ as a band index and $\vartheta_{\vec{k}_l} = \arctan(k_l^y/k_l^x)$. For BGr, $\vec{\chi}_{T,\xi} = (\chi^A_{T,\xi}, \chi^B_{T,\xi})$ is the two component vector formed from the lower two components of the normalized eigenvectors of Hamiltonian (S.2) (i.e. wavefunction components on the layer of the BGr closest to the Gr). The Bloch wavefunctions, $\phi^i_{l,\xi}$, on the $i = A/B$ sublattice of a Gr layer, or the lower layer of the BGr are written,

$$\phi^{i=A/B}_{l,\xi}(\vec{r}, z) = \frac{3^{1/4} a}{\sqrt{2} L} \sum_{\vec{R}_l} e^{i(\vec{K}_{l,\xi} + \vec{k}_l) \cdot (\vec{R}_l + \vec{\tau}^i_l)} \psi_l(\vec{r} - \vec{R}_l - \vec{\tau}^i_l, z), \quad (S.4)$$

where $\psi_l$ are the carbon P$_z$ orbitals and $L^2$ is the area of the flake. For the bottom layer, $\vec{R}_B$ are the lattice positions, $\vec{\tau}^A_B = (0, a/\sqrt{3})$ and $\vec{\tau}^B_B = (0, -a/\sqrt{3})$ are the nearest neighbor vectors, $a$ is the lattice constant, and $\vec{K}_{B,\xi} = (\xi\frac{4\pi}{3a}, 0)$ is the valley center. Corresponding quantities on the top layer are rotated by the misalignment angle $\theta$. To evaluate the overlap integrals in Eq. (S.3) we rewrite Bloch wavefunctions (S.4) using the in-plane Fourier transform of the P$_z$ orbitals, $\hat{\psi}_l(\vec{q}, z)$,

$$\phi^i_{l,\xi}(\vec{r}) = \frac{4 \times 3^{1/4} \pi}{a\sqrt{6} L} \sum_{\vec{g}_l} \hat{\psi}_l(\vec{K}_{l,\xi} + \vec{g}_l + \vec{k}_l, z) e^{i(\vec{K}_{l,\xi} + \vec{g}_l + \vec{k}_l) \cdot \vec{r}} e^{-i\vec{g}_l \cdot \vec{\tau}^i_l}, \quad (S.5)$$

where $\vec{g}_l$ are the reciprocal lattice vectors of graphene layer $l$. Using this we obtain,

$$t\langle \phi^i_T | \phi^j_B \rangle = \sum_{\vec{g}_T, \vec{g}_B} \Gamma_{\vec{g}_T, \vec{g}_B} \delta(\vec{g}_T + \vec{K}_{T,\xi} - \vec{g}_B - \vec{K}_{B,\xi} + \vec{k}_T - \vec{k}_B) e^{i\vec{g}_T \cdot \vec{\tau}^i_T} e^{-i\vec{g}_B \cdot \vec{\tau}^j_B}$$

$$\Gamma_{\vec{g}_T, \vec{g}_B} \approx \frac{32\pi^4 t}{\sqrt{3} a^2 L^2} \int dz\, \hat{\psi}^*_T(\vec{K}_{T,\xi} + \vec{g}_T, z) \hat{\psi}_B(\vec{K}_{B,\xi} + \vec{g}_B, z) \quad (S.6)$$

The approximation in the second line used the fact that $|\vec{g}_T|, |\vec{g}_B| \approx a^{-1}$, where as only states with $\vec{k}_l \ll a^{-1}$ contribute to the tunneling process. For the same reason, we must also select $\vec{g}_T = \vec{g}_B$ (up to rotation by $\theta$) in the sum in the first line so that the Dirac-delta function can be satisfied. Next, we use the fact that the separation between the Gr/BGr flakes is considerably larger than the Bohr radius of the carbon P$_z$ orbitals. Because of this, $\hat{\psi}^l(\vec{q}, z)$ decays rapidly for $|\vec{q}| \gg |\vec{K}|$ at distances $z$ comparable to the separation between layers, so that we retain only those three terms for which $|\vec{K}_{B,\xi} + \vec{g}_B| = |\vec{K}_{B,\xi}|$. Using this in Eq. S.3 we obtain,

$$M^\xi_{\vec{k}_T, \vec{k}_B} = \Gamma_{\vec{0}, \vec{0}} \sum_{n=0,1,2} \delta(\vec{k}_T + \Delta\vec{K}_{n,\xi} + \Delta\vec{p} - \vec{k}_B) g_T(\vec{k}_T) g_B(\vec{k}_B), \quad (S.7)$$

$$g_l(\vec{k}_l) = (\chi^A_{l,\xi} + \chi^B_{l,\xi} e^{-i\xi\frac{2\pi n}{3}}),$$

where,

$$\Delta\vec{K}_{n,\xi} = \xi R_{\frac{2\pi}{3}n}[R_\theta - 1]\vec{K}_{B,\xi}, \quad (S.8)$$

is written using the anticlockwise rotation matrix $R_\phi$ and represents the k-space shift between the six ($\xi=\pm, n=0,1,2$) corners of the first Brillouin Zones on the two layers caused by their misalignment. Also, the term

$$\Delta\vec{p} = ed_{\text{hBN}} \vec{l}_z \times \vec{B}_\| \quad (S.9)$$

has been included into the Dirac-delta function to account for the Lorentz boost (*19, 20*) produced by the magnetic field. Here $e$ is the electron charge, and $d_\text{hBN} = 1.4$nm for the Gr/hBN/Gr device or $d_\text{hBN} = 2$nm for the Gr/hBN/BGr device is the thickness of the hBN layers separating the Gr/BGr electrodes.

We treat the tunneling perturbatively to obtain the expression for the tunneling current (*42, 43*),

$$I = \frac{g_s e L^4}{(2\pi)^5} \sum_{s_B, s_T, \xi} \int d\vec{k}_B d\vec{k}_T d\epsilon \, |M^\xi_{\vec{k}_T, \vec{k}_B}|^2 A(\vec{k}_B, \epsilon) A(\vec{k}_T, \epsilon - u)[f_B(\epsilon) - f_T(\epsilon - u)]. \tag{S.10}$$

Here $g_s = 2$ is the spin degeneracy, $f_{B/T}$ are the Fermi occupancy factors, and $u$ is the band offset between the two layers (see section S2). Also

$$A(\vec{k}_l, \epsilon) = \frac{2\gamma_l}{(\epsilon - \epsilon_l)^2 + \gamma_l^2} \tag{S.11}$$

is the spectral function, with $\epsilon_l$ the appropriate eigenvalue of Hamiltonian (S.1) or (S.2), $s_{B/T}$ is the band index (summed over all eigenvalues), and $\gamma_l$ is the inelastic electron life-time.

For the Gr/hBN/Gr device we calculate the inelastic electron life-time, $\gamma_l = \text{Im}\Sigma_\text{e-e} + \text{Im}\Sigma_\text{e-ph}$, using the imaginary parts of the retarded self-energies of the RPA-screened Coulomb interaction, calculated following the prescriptions in Refs. (*44, 45*), and the emission of phonons from the highest optical branches at the $K$ (*46, 47*) and $\Gamma$-points (*46, 48*). However, we found that the best match between theory and experiment (used throughout) was obtained by doubling the level broadening calculated using the above-described method. We attribute this to the fact that the inelastic relaxation of hot carriers may involve cold electrons in both the collector and emitter. For the Gr/hBN/BGr device we simply take $\gamma_{T/B} = 0.03|\epsilon_{T/B} - \mu_{T/B}| + 2$ meV, which was chosen for a good fit to the experiment, and to mimic the expected increase in level broadening away from the Fermi energy, $\mu_{T/B}$, in the appropriate layer.

To simplify the numerical evaluation of Eq. (S.10) (except in Fig. 4 of the main text where numerical integration over $\epsilon$ is used), we use the Dirac-delta function in Eq. (S.7) to evaluate the integral over $\vec{k}_T$ and approximate $\gamma_{T/B} \ll eV_b$ in the integral over $\epsilon$ to obtain,

$$I = \frac{g_s e L^6 \Gamma^2_{\vec{0},\vec{0}}}{(2\pi)^3} \sum_{s_B, s_T, \xi, n} \int d\vec{k}_B \frac{\text{Im}}{\pi} \frac{|g_T(\vec{k}_B - \Delta\vec{K}_{n,\xi} - \Delta\vec{p})g_B(\vec{k}_B)|^2}{(\epsilon_T + u - \epsilon_B) - i(\gamma_B + \gamma_T)}[f_B(\epsilon_B) - f_T(\epsilon_B - u)]. \tag{S.12}$$

The factors $g_l(\vec{k}_l)$ in Eq. (S.12) encode the interference of wavefunction amplitudes on the $A/B$ sublattices. This interference is shown schematically in Fig. 1 (C,G) of the main text, which displays the real part of the wavefunction overlapped with a plane wave, $e^{-i\vec{K}\cdot\vec{r}}(\chi^A_{l,\xi}\phi^A_{l,+} + \chi^B_{l,\xi}\phi^B_{l,+})$, $\chi^A_{l,\xi} = 1/\sqrt{2}$, for either $\chi^B_{l,\xi} = 1/\sqrt{2}$ (constructive interference), or $\chi^B_{l,\xi} = -1/\sqrt{2}$ (destructive interference). Fig. 1 (D,H) of the main text displays the interference of spherical waves centered on the $A/B$ sites in (C,G) as a function of distance from the graphene layer.

## 4 Electrostatic model for Gr/hBN/Gr and Gr/hBN/BGr devices

We describe the electrostatic properties of the devices using a three-plate capacitor model, consisting of the doped n-Si back gate and the two Gr/BGr electrodes. Based on this model (*23*) we derive the following pair of simultaneous equations,

$$eV_b = \frac{e^2}{\epsilon_{\text{hBN}}\epsilon_0}d_{\text{hBN}}n_T + \mu_T - \mu_B,$$
$$eV_g = \frac{-e^2}{\epsilon_{\text{SiO}}\epsilon_0}d_{\text{SiO}}(n_B + n_T) - \mu_B. \tag{S.13}$$

Here $\epsilon_0$ is the vacuum permittivity, we use $\epsilon_{\text{hBN}} \approx 3.2$ and $\epsilon_{\text{SiO}} \approx 3.9$ for the dielectric constants of hexagonal boron nitride and silicon oxide, $d_{\text{SiO}} \approx 300$ nm is the oxide thickness to the n-Si backgate, $n_{T/B}$ and $\mu_{T/B}$ is the carrier density and chemical potential on the top/bottom electrode, and the band offset is obtained using $u = e^2 d_{\text{hBN}} n_T/(\epsilon_{\text{hBN}}\epsilon_0)$.

For the Gr/hBN/Gr device at zero magnetic field, we use

$$n_{T/B} = \text{sign}(\mu_{T/B})\frac{\mu_{T/B}^2}{\pi v^2},$$

and solve equations (S.13) numerically for $\mu_B$ and $\mu_T$. When the Gr/hBN/Gr device is placed in a rotating magnetic field (Fig. 3 H of the main text), we notice the presence of a feature at $V_b \approx 0.26$V (corresponding to electrostatic conditions for which $\mu_B \approx 0$) which varies with the orientation of the magnetic field, $\alpha$, with a periodicity of $\pi$. This is attributed to the presence of a finite out-of-plane component of the magnetic field $B_\perp \propto \sin(\alpha + \delta\alpha)$ created by a slight misalignment of the device in the sample holder. We incorporated this into the theoretical model by including the effect of Landau level formation on the carrier density. For a pristine graphene flake, a perpendicular magnetic field quenches the density of states into a series of Dirac delta-function like peaks at the energy of each Landau level (*49*),

$$\epsilon_m = \text{sign}(m)v\sqrt{2|meB_\perp|}.$$

Here we account for a small Gaussian broadening, $\sigma$, in each Landau level so that the carrier density reads,

$$n(\mu_{T/B}) = \frac{g_s g_v B_\perp}{\phi_0} \sum_{m=-\infty}^{\infty} \int_0^{\mu_{T/B}} d\epsilon \frac{1}{\sqrt{2\pi}\sigma} e^{-(\epsilon-\epsilon_m)^2/(2\sigma^2)}$$
$$= \frac{g_s g_v B_\perp}{2\phi_0} \sum_{m=-\infty}^{\infty} \left[\text{erf}\left(\frac{\epsilon_m}{\sqrt{2}\sigma}\right) - \text{erf}\left(\frac{\epsilon_m - \mu_{T/B}}{\sqrt{2}\sigma}\right)\right],$$

where $\frac{g_s g_v B_\perp}{\phi_0}$ is the capacity of each Landau level to hold electrons, $\phi_0 = h/|e|$ is the flux quanta, $g_s g_v = 4$ the spin-valley degeneracy, and we use $\sigma = 5$ meV, a maximum perpendicular field of $0.5$T (corresponding to a missorientation of the magnetic field of approximate $1°$), and $\delta\alpha = -40°$ to model the experimental data.

For the Gr/hBN/BGr device we use $n_B = \text{sign}(\mu_B)\mu_B^2/(\pi v^2)$ for the bottom electrode, but must calculate $n_T = n_T(\mu_T, \Delta)$ numerically from Hamiltonian (S.2) as a function of $\mu_T$ and $\Delta$ (setting $v_3 = v_4 = 0$ here for simplicity). The bandgap is obtained self-consistently using the three plate capacitor model, where we also account for the screening of the gap caused by the redistribution of the the BGr wavefunction between the two BGr layers in response to the gap itself. Approximating $|\Delta| \ll \gamma_1$, and using Ref. (*40*) we find,

$$\Delta = \frac{e^2 d'}{\epsilon_0} \frac{n_T}{2} \left(1 + \frac{\Lambda v^2 \pi |n_T|}{\gamma_1^2} - \frac{\Lambda}{2} \log\left(\frac{v^2 \pi |n_T|}{\gamma_1^2}\right)\right)^{-1} \quad (S.14)$$

where $d' = 3.4\,\text{Å}$ is the interlayer distance within the BGr, and $\Lambda = e^2 \gamma_1 d'/(2\pi v^2 \epsilon_0) \approx 1$ controls the effectiveness of the screening (*40*). Note that for $\Lambda = 0$ the electron density $n_T$ is equally distributed between the layers of BGr so that the expression (S.14) is reduced to $\Delta = \frac{e^2 d'}{\epsilon_0} \frac{n_T}{2}$. We take $\Lambda = 1$ and solve equations (S.13) and (S.14) numerically for $\mu_B$, $\mu_T$, and $\Delta$.

## 5  Calculated $dI/dV_b$ maps of Gr/hBN/BGr devices

To mimic the supposed presence of two regions with different misalignment angles with in the active area of the Gr/hBN/BGr device, we modeled the total tunneling current by summing the currents obtained using misalignment angles $\theta = 0.5°$ and $\theta = -3.1°$. The I-V characteristics produced by these two angles are displayed separately in Figs. S2 and S3, which also highlights the various band alignments responsible for characteristic features. Notice that there is a window of small voltages $|V_b| \lesssim 0.4\,\text{V}$, $|V_g| \lesssim 50\,\text{V}$ for which the modeled tunneling current is produced entirely by the $\theta = 0.5°$ region of the device. It is this range of voltages which produces the best agreement between theoretically calculated and experiment measured currents.

## 6  $dG/dV_b$ map at $V_g = -45\,\text{V}$

Figure S4 displays an additional example of a $dG/dV_b$ map for the Gr/hBN/BGr device discussed in the main text, measured or calculated for gate voltage $V_g = -45V$.

## 7  Detection of the effects of chirality

In equation (S.12), the sublattice composition of the electron wavefunction is encoded in the "chirality factors", $|g_{T/B}|^2$. A color map of these factors is overlayed on the bandstructures of Gr and BGr in Fig. S5 A. In this section we show further proof of our ability to identify features in the $dG/dV_b$ maps caused by these factors by comparing the experimental data with the calculated version using either the full chirality factors, or by setting them to a constant $g_{B/T} = 1$.
This analysis is simplest for the Gr/hBN/Gr device. Here any observed asymmetry under reflecting $\alpha \to -\alpha$ in the $dG/dV_b$ maps must be caused by the chirality factors, as these are the only factors in Eq. (S.12) which are sensitive to the direction (rather than just the magnitude) of $\Delta \vec{K}_{n,\xi} + \Delta \vec{p}$. In particular, when the bands are aligned as per Fig. 2 B of the main text, the rotation of $\Delta \vec{p}$ systematically pushes the resonant tunneling states to regions of the Dirac cones for which the chirality factors enhance or suppress the tunneling. As a result, the corresponding feature in the $dG/dV_b$ map (found at $V_b \approx 0.8V$ in Fig. 3 H,I of the main text) displays a weak $\alpha \to -\alpha$ asymmetry. In contrast, when the chirality factors are set to a constant value, Fig. S5 B, there is perfect $\alpha \to -\alpha$ symmetry.

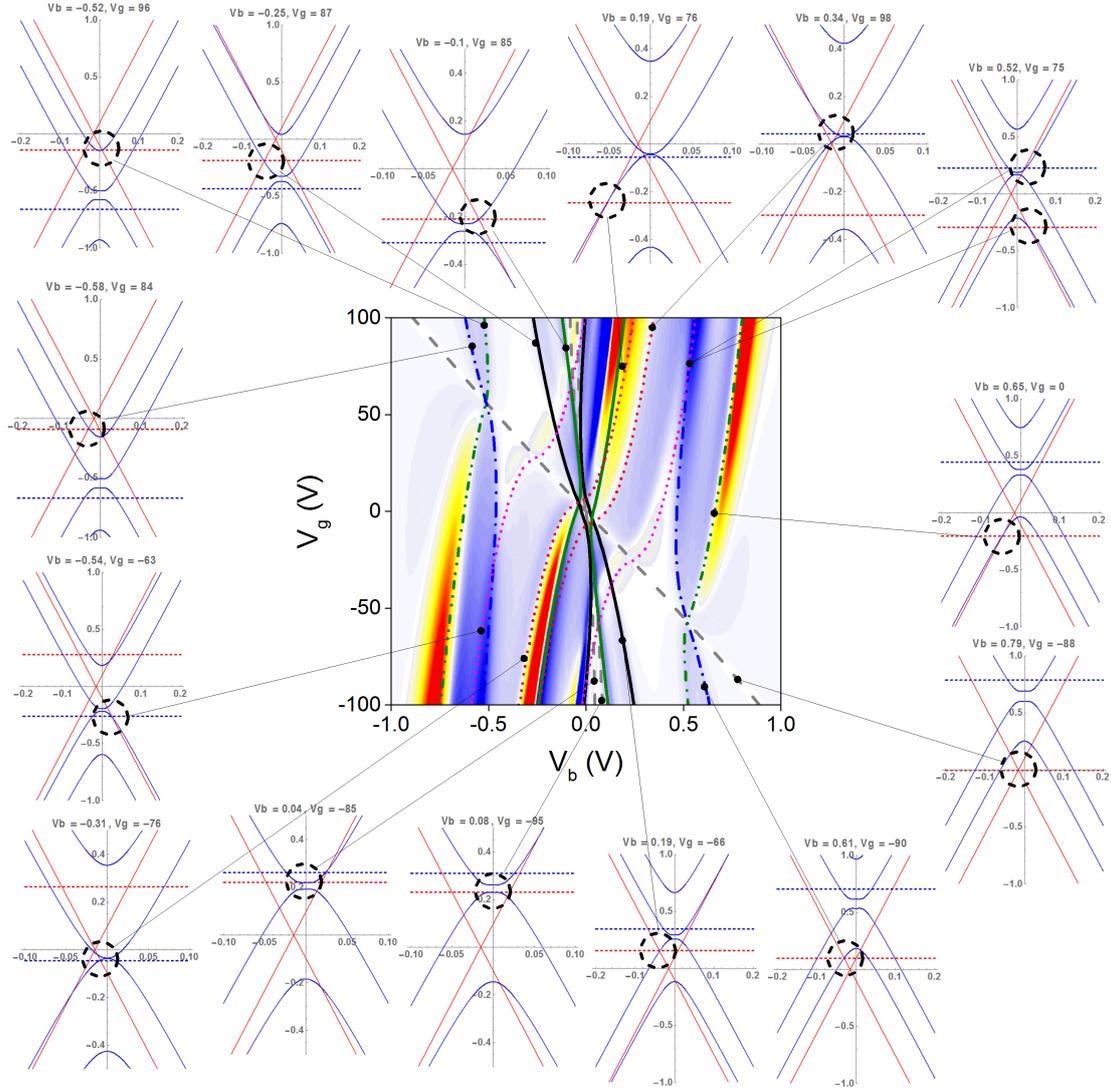

Figure S2: Differential tunnel conductance calculated for Gr/5hBN/BGr device with misalignment angle 0.5°. Various resonances are highlighted on the map and schematically plotted on the cartoons. In the cartoons monolayer graphene is shown in red, bilayer graphene - in blue, their respective Fermi levels are shown as red and blue horizontal dashed lines.

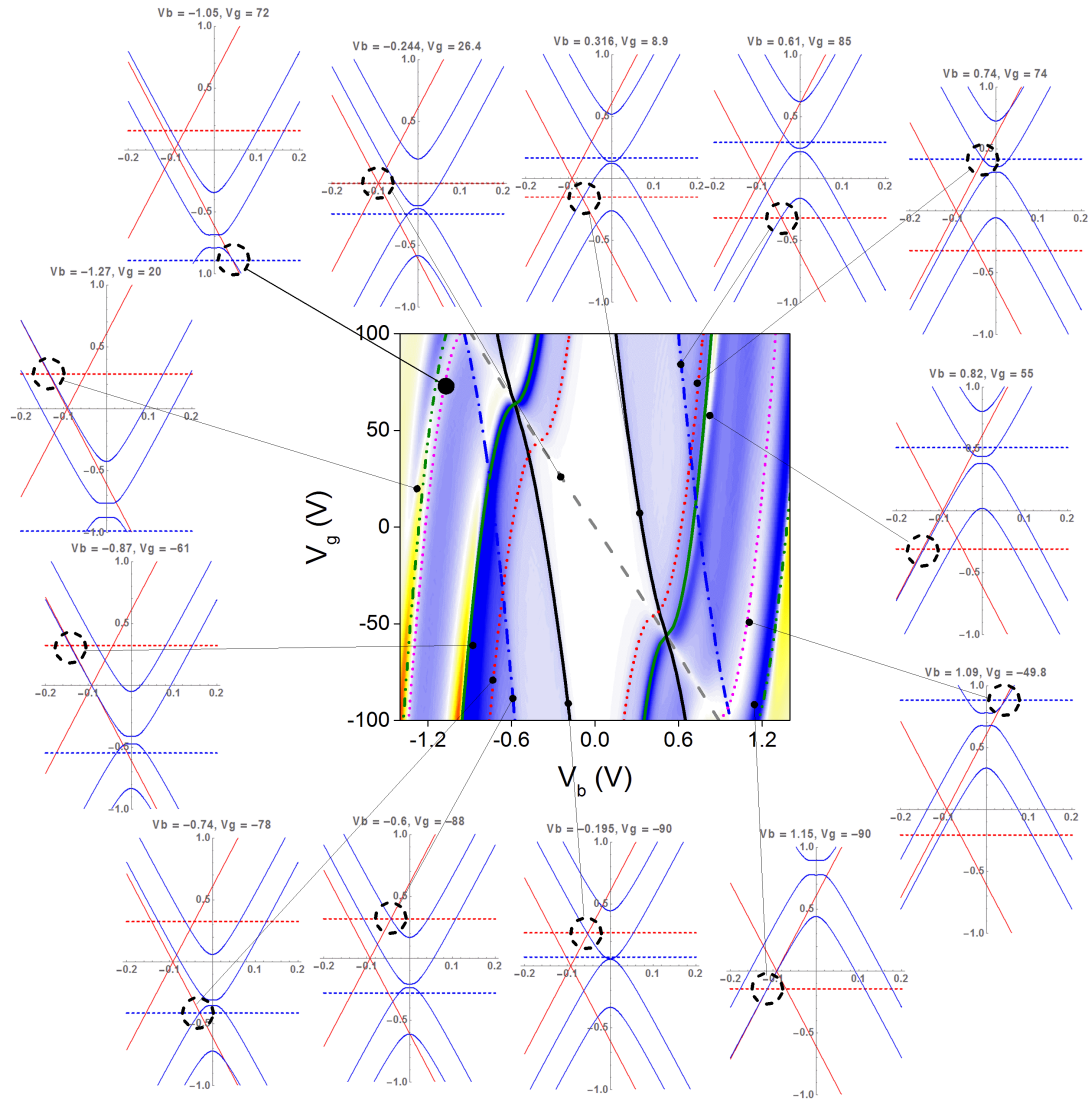

Figure S3: Differential tunnel conductance calculated for Gr/5hBN/BGr device with misalignment angle -3.1°. Various resonances are highlighted on the map and schematically plotted on the cartoons. In the cartoons monolayer graphene is shown in red, bilayer graphene - in blue, their respective Fermi levels are shown as red and blue horizontal dashed lines.

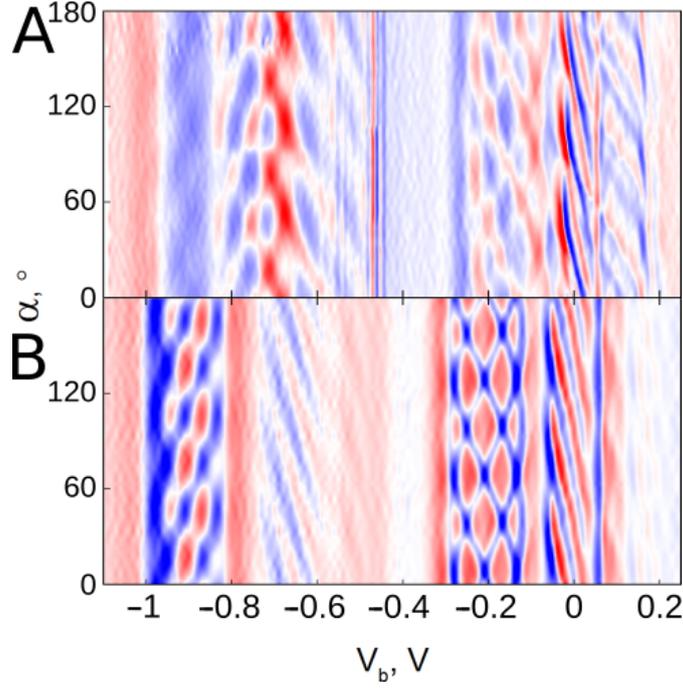

Figure S4: The $dG/dV_b$ map for the Gr/hBN/BGr device discussed in the main text, either measured (A) or calculated (B) for gate voltage $V_g = -45V$ and $B = 30$ T.

For the Gr/hBN/BGr device the $\alpha \rightarrow -\alpha$ asymmetry is far more striking. However, for BGr there is a non-negligible trigonal warping of the bandstructure, which can produce a $\alpha \rightarrow -\alpha$ asymmetry in a similar manner to the chirality factors. To discriminate between these two effects, Fig. S5 (C-E) and (F-H) display $dG/dV_b$ maps calculated for gate voltages $V_g = 60V$ and $V_g = -45V$, using either (i) constant chirality factors, (ii) the absence of trigonal warping, or (iii) both constant chirality factors and no trigonal warping. In particular, the feature displaying very strong $\alpha \rightarrow -\alpha$ asymmetry, found between $V_b = 0$ and $0.25$ V at $V_g = 60V$ (Fig. 3 K,L of the main text), occurs at gate and bias voltages which are approximately inverted compared to the feature between $V_b = -0.25$ and $-0.05$ V at $V_g = -45V$ (Fig. S4). Thus, while the feature in the $V_g = 60V$ map is produced by the onset of resonant tunneling in the valence band, the feature in the $V_g = -45V$ map is produced by a similar resonant onset but in the conduction band. This results in the trigonal warping breaking the $\alpha \rightarrow -\alpha$ symmetry in the opposite manner for the two features, which is highlighted by the dashed arrows in Figs. S5 C,F. In contrast, the chirality factors break the $\alpha \rightarrow -\alpha$ symmetry in the same manner in both cases (see dashed arrows in Figs. S5 D,G), resulting in a competition between the two effects for the discussed feature at $V_g = -45V$. By comparing these images with the corresponding experimentally measured maps, it is clear that both effects are resolved and that the symmetry breaking effect of chirality is the strongest (in Fig. S4 A the $\alpha \rightarrow -\alpha$ symmetry is broken in the manner determined by the chirality factors rather than trigonal warping).

## 8 Maximum obtainable valley polarization

Here we study the valley polarization for a device with two monolayer graphene electrodes, since this crystal has a steeper energy dispersion than bilayer graphene and, consequently, a given Lorentz boost of momentum can result in a greater shift in the resonant tunneling conditions and a correspondingly greater valley polarization. For this system valley polarization typically increases with the dimensionless ratio $x = v|\Delta\vec{p}|/\gamma$, where $\gamma = \langle \gamma_T + \gamma_B \rangle$ is the typical lifetime broadening for the states involved in tunneling. As a result, one strategy to maximize the valley polarization is to tune the device parameters so that features in the tunneling current occur at low bias voltages where $\gamma$ is expected to be reduced.

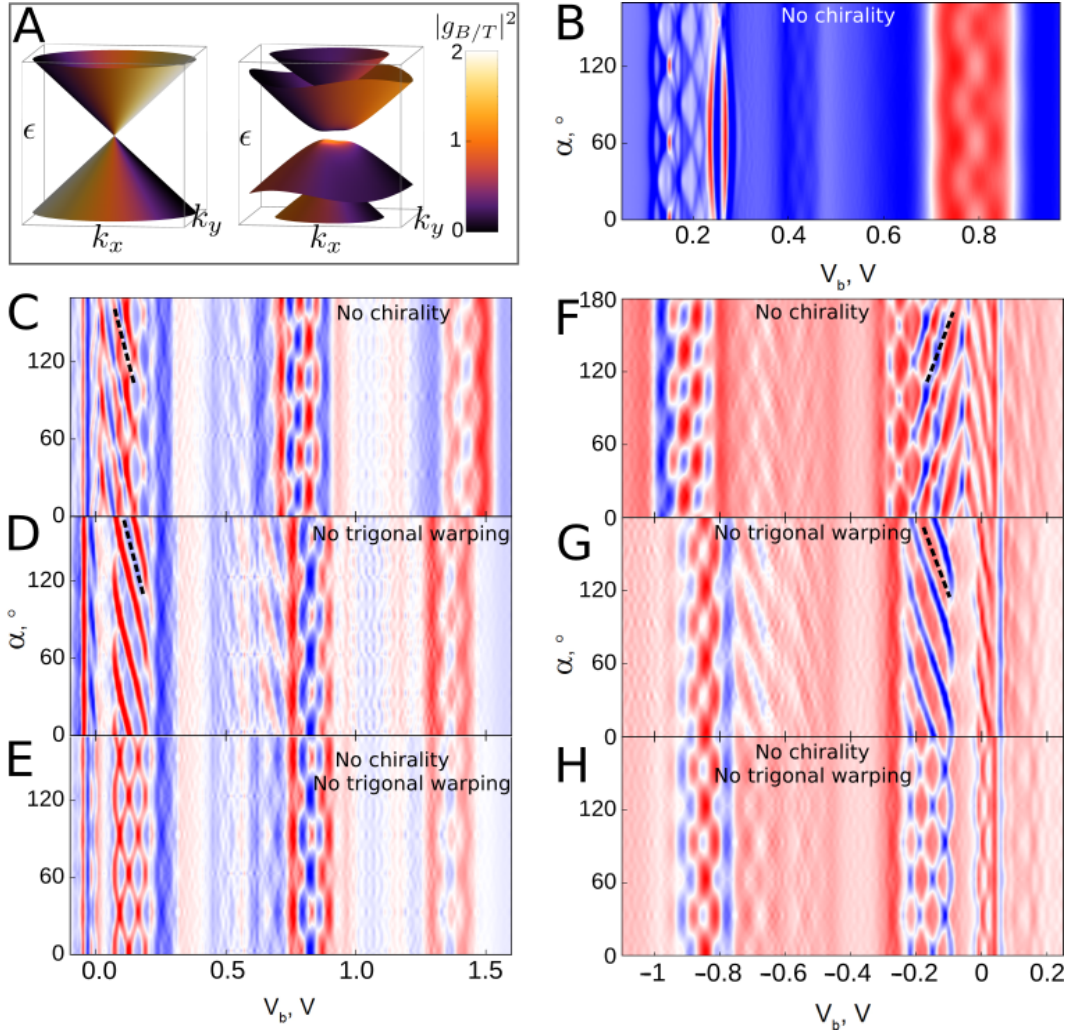

Figure S5: The effects of chirality on the tunneling current (A) A color map of the chirality factors of Gr (left) and BGr (right) overlayed on their bandstructures. This image was produced for the $\xi = 1$, $j = 0$ Brillouin zone corner using a gap of $\Delta = +0.1\,\text{eV}$ for BGr. Corresponding images for different Brillouin zone corners can be obtained by $60°$ rotations. (B) The $dG/dV_b$ map for Gr/hBN/Gr calculated using the same parameters as Fig. 3 I of the main text, but setting $|g_{B/T}| = 1$. (C-E) The $dG/dV_b$ map for the Gr/hBN/BGr device at $V_g = 60V$ calculated using the same parameters as Fig. 3 L of the main text, but either setting $|g_{B/T}| = 1$ (C), $v_3 = v_4 = 0$ (D), or $|g_{B/T}| = 1$ and $v_3 = v_4 = 0$ (E). (F-H) The same as (C-E) except calculated for $V_g = -45V$ using the same parameters as Fig. S4 B.

To estimate the maximum obtainable valley polarization for a fixed ratio $x$, we retain only the Lorentzian factor from Eq. (S.12) and focus on conditions expected to maximize the valley polarization, so that we take the tunneling current generated in the vicinity of Brillouin zone corner $(n, \xi)$ as

$$I_{n,\xi} \sim \frac{\gamma}{\Delta\epsilon_{n,\xi}^2 + \gamma^2}. \tag{S.15}$$

Here $\Delta\epsilon_{n,\xi} = v|\vec{k}_{n,\xi} + \Delta\vec{p}|$, and $\vec{k}_{n,\xi} = \xi R_{2\pi n/3}\vec{k}_{n=0,\xi=1}$ is the wavevector for a small patch of k-space assumed to be responsible for the maximum valley polarization ($R_\phi$ is the rotation matrix). Then, valley polarization is,

$$P = \frac{I_K - I_{K'}}{I_K - I_{K'}} = \frac{2x^3 X^3 \sin(3\phi)}{(x^2+1)^3 + X^6 + 3X^4 + 3(x^2+1)X^2}. \tag{S.16}$$

where $I_K = \sum_{n=0,1,2} I_{n,\xi=1}$ and $I_{K'} = \sum_{n=0,1,2} I_{n,\xi=-1}$ are the currents produced in the $K$ and $K'$ valleys, $X = v|\vec{k}_{0,1}|/\gamma$, $\phi$ is the angle between $\Delta\vec{p}$ and $\vec{k}_{0,1}$. An upper limit on the valley polarization,

$$P_{\max} = \frac{x^3}{\sqrt{x^2+1}(x^2+4)}. \tag{S.17}$$

is obtained for $X = \sqrt{1+x^2}$ and $\phi = -\pi/2$.

In Fig. S6, $P_{\max}$ is shown using a solid red line. This is compared to the valley polarizations, calculated numerically using Eq. (S.12) for the Gr/hBN/Gr device discussed in the main text with $B = 30\,\text{T}$ (and using $\gamma_B + \gamma_T$ to vary $x$). Here blue dots are used for $V_g = -20\,\text{V}$, $V_b \approx 0.13\,\text{V}$ and red dots for $V_g = -20\,\text{V}$, $V_b \approx 0.74\,\text{V}$, which approximately correspond to band alignments displayed in the blue and purple panels on the right (the exact value of $V_b$ was allowed to vary slightly to find a local maxima in $P$ but $V_g$ was fixed to the value used in Fig 3K of the main text). For these conditions the obtained valley polarization is lower than the maximum $P_{\max}$ but nevertheless saturates at $100\%$ valley polarization as $x \to \infty$. To obtain valley polarization approaching $P_{\max}$ the device parameters must be carefully tuned. For example, for a device with misalignment angle $\theta = 0.2°$ (red dots) the Lorentz boost from a 30 T magnetic field fully compensates for the misalignment angle for one of the Brillouin zone corners only, to produce strong valley polarization (e.g $P \approx 85\%$ for $x = 7$).

## 9 Comparison of aligned and non-aligned tunnel transistors

Fig. S7 compares the characteristics of two Gr/hBN/Gr tunneling transistors: the one where two graphene layers are rotationally misaligned by a large angle of $> 10°$ (Fig. S7 E), and the one with two graphenes crystallographically aligned within $\approx 2°$ (Fig. S7 F). In both cases graphene layers were misaligned with respect to both hBN substrate and spacer.

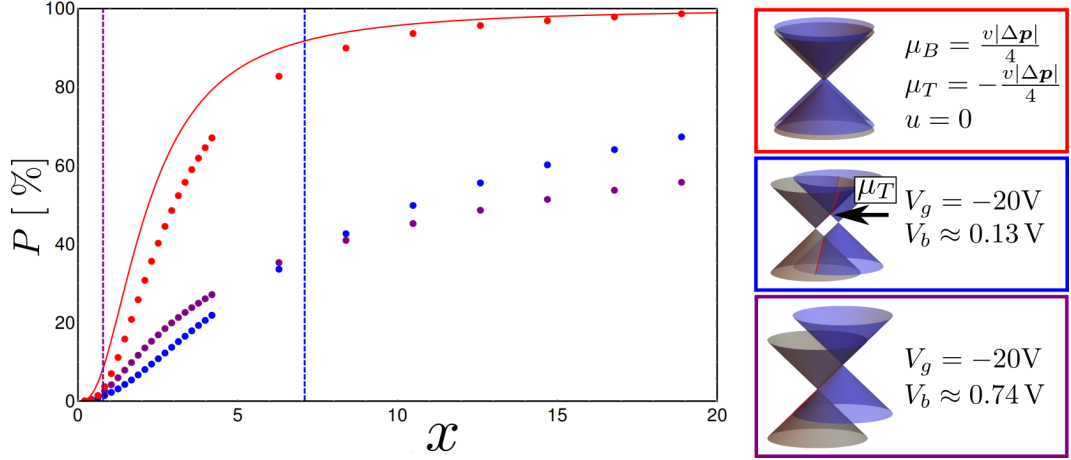

Figure S6: The valley polarization, $P$, as a function of $x = v|\Delta \vec{p}|/\gamma$, showing $P_{\max}$ (solid red line), and the numerically calculated valley polarization (blue, purple and red dots) for the three situations displayed in the blue, purple and red boxed panels on the right (see text). For comparison, vertical blue and purple dashed lines display the values of $x$ calculated for $V_b = 0.13\,\text{V}$ or $V_b = 0.74\,\text{V}$ with $V_g = -20\,\text{V}$ and $B = 30\,\text{T}$ using the calculation of the electron lifetime described in supplementary section S3.

Two features are very similar for both devices: the two intersecting blue lines showing the condition of the Fermi level passing through the Dirac point in one of the layers Fig. S7 B. Here the vanishing density of states leads to a reduced tunnel conductance. This feature does not rely on the momentum conservation and thus it appears in both aligned and non-aligned devices.

The most important and the most obvious difference between aligned and non-aligned tunnel transistors is the absence of negative differential conductance (NDC) in non-aligned case. While for aligned sample the NDC regions are prominent (see Fig 2C of the main text), we never observed NDC in nonaligned samples. Apart from the main resonance the onsets of four other resonances are noticeable (i.e. yellow lines on Fig. S7, F). These mark the transition from non-resonant to resonant tunneling (schematically shown on Fig. S7 C). Another strong difference is the presence of phonon-assisted tunneling seen as vertical (independent of $V_g$) white stripes on the conductance map of a non-aligned sample (marked by green line on Fig. S7 E). This is because in non-aligned samples the large mismatch of in-plane momentum between the top and bottom graphene layers has to be compensated for tunneling electrons. Emitting a phonon of specific energy $\hbar\omega$ provides the matching momentum for the tunneling electron and thus the momentum conservation is relaxed (Fig. S7 A). More details about phonon assisted tunneling in non-aligned samples can be found elsewhere (*24, 25, 26*).

## 10 Devices with nearly-perfectly aligned graphene electrodes

Fig. S8 presents a conductance map for a device with nearly perfectly aligned graphene electrodes, Fig. S8 A (alignment better than $0.1°$). It resembles closely the theoretically calculated conductance map for a device with perfect alignment, Fig. S8 B. Small differences are due to small, unintentional doping of one of the graphene electrodes. One can see that different resonances described above are collapsed into a single one, which is located close to $V_b = 0$ V. Unfortunately devices with zero misalignment do not offer a possibility to arrange valley or pseudospin polarization in magnetic field, Fig S9.

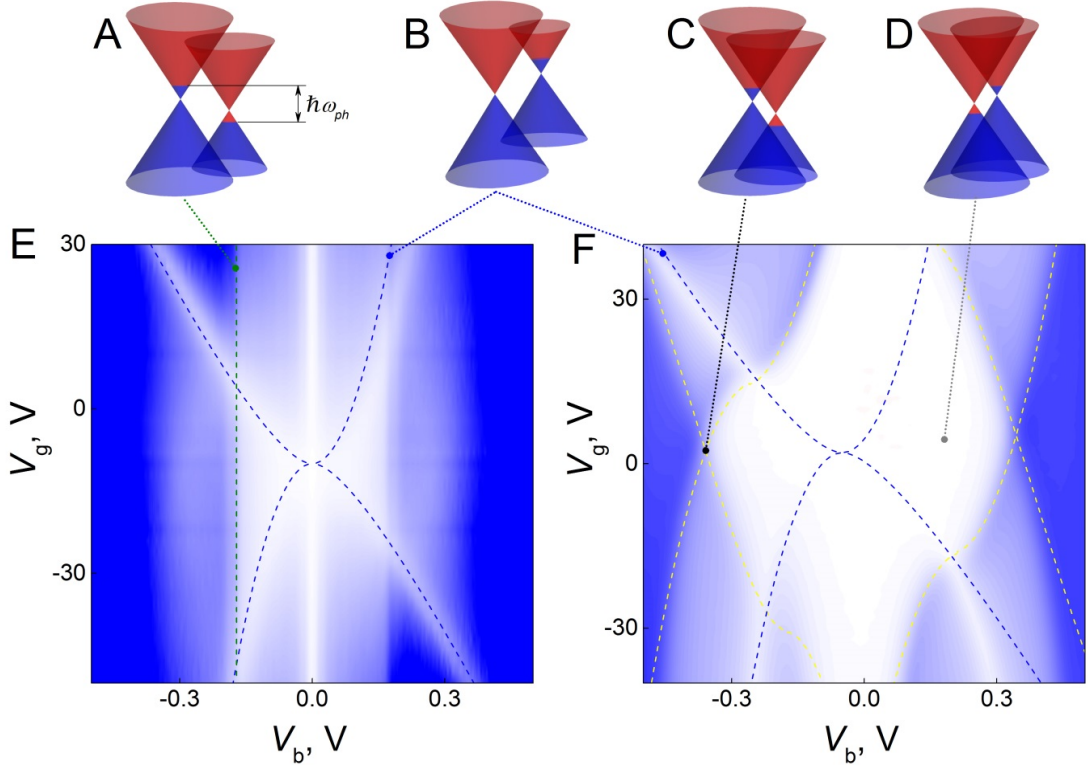

Figure S7: Tunnelling conductance maps ($dI/dV_b$ vs bias voltage, $V_b$, and gate voltage, $V_g$) for non-aligned (E) and aligned (F) tunnel transistors, measured at 2K. Color scheme is the same for both panels: white to blue is $0\,\mu$S to $+5\,\mu$S. The lines on the conductance maps indicate resonances, which correspond to specific conditions of the position of the Fermi levels in the two graphene electrodes. The green line on (E) corresponds to the situation when the bias is large enough to allow tunneling with emission of a phonon (situation schematically described in panel (A)). Blue lines on (E) and (F) correspond to the Fermi level in one of the electrodes passing through the Dirac point (situation schematically described in panel (B)). Yellow lines on (F) correspond to the Fermi level in one of the electrodes passing through crossing points between the Dirac cones (situation schematically described in panel (C)). Note, that at lower bias voltages, when the Fermi level doesn't pass through the crossing between the Dirac cones (situation schematically described on panel (D)), the conductivity is strongly suppressed.

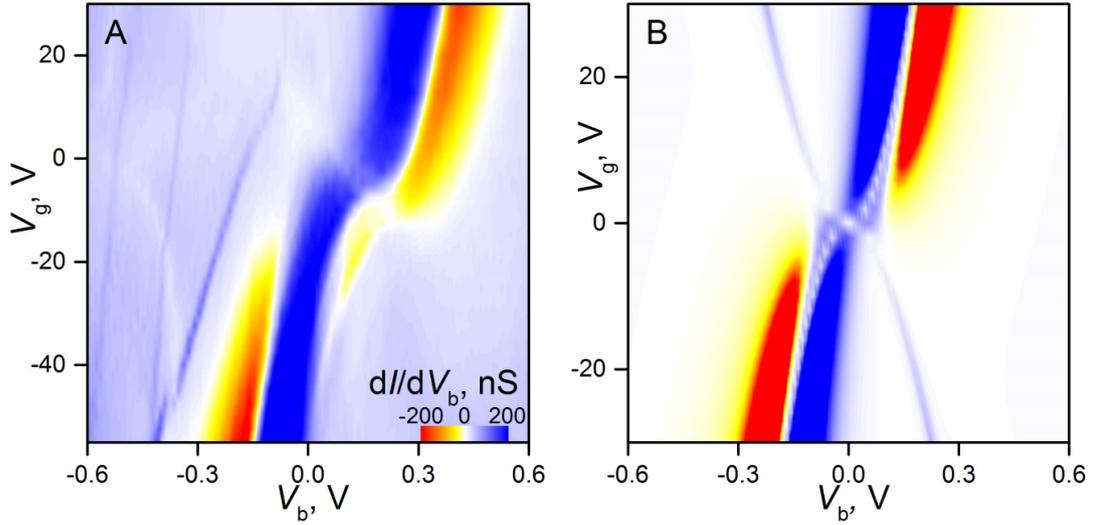

Figure S8: Tunnelling conductance maps ($dI/dV$ vs bias voltage, $V_b$, and gate voltage, $V_g$) for a device with nearly perfect alignment between the graphene electrodes. (A) experiment; (B) theory for perfectly aligned graphene electrodes (color scale : arbitrary units).

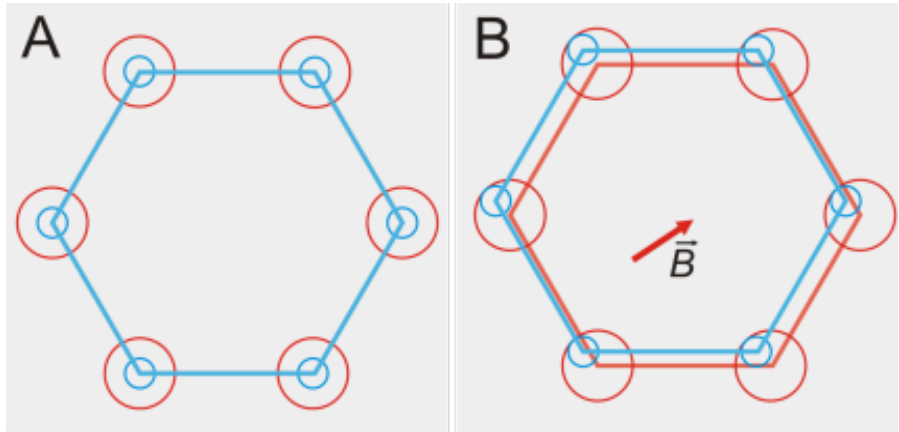

Figure S9: A schematics of magnetotunneling for the case of perfectly aligned graphene electrodes. (A) Cartoon representing the Brillouin zones of the emitter (blue) and collector (red) graphene electrodes. Circles denote the cuts through graphene bandstructure at the Fermi energy in the two graphene layers. Different size of the circles for emitter and collector are due to the applied large gate voltage, this introduces the momentum mismatch for tunneling electrons. (B) As (A) but with magnetic field applied parallel to graphene layers. The Lorentz force leads to an additional momentum acquired by electrons when they tunnel from emitter to collector graphene electrode, this is seen as a shift of the Brillouin zones with respect to each other by the vector. In case of perfectly aligned graphene electrodes this brings all six corners of the Brillouin zone into resonance (momentum conservation) simultaneously. Thus, the valley polarization is absent in the case of $0°$ misalignment.